\documentstyle[epsf,12pt]{ioplppt}
\begin{document}

\title{X-ray photoelectron 3s spectra of transition metal oxides}

\author{V R Galakhov\dag, S Uhlenbrock\ddag, S Bartkowski\ddag,
A V Postnikov\ddag,\\
M Neumann\ddag, L D Finkelstein\dag, E Z Kurmaev\dag, A A Samokhvalov\dag\\
and L I Leonyuk\S}

\address{\dag\ Institute of Metal Physics, Russian Academy of Sciences,
Ural Division \\ 620219 Ekaterinburg GSP-170, Russia}
\address{\ddag\ University of Osnabr\"uck, Fachbereich Physik,
D-49069 Osnabr\"uck, Germany}
\address{\S\ Department of Crystallography, Faculty of Geology,
Moscow State University, 119899 Moscow, Russia}

\begin{abstract}
We present  metal 3s  x-ray photoelectron spectra of different
transition metal oxides.
The values of the 3s exchange splitting for the 3d metal oxides are
given as a function of the 3d electron number in the ground state.
The spectra were analysed using the simple two-configuration
model of the interatomic configuration mixing.
The change of the 3s spectra is ascribed to the change of the
charge-transfer energy.
\end{abstract}

\pacs{79.60, 71.70}
\maketitle

\section{Introduction}
\label{sec:intro}

The spectral splitting of 3s core-level x-ray photoemission spectra
(XPS) in transition metals and their compounds originates from the
exchange coupling between the 3s hole and the 3d electrons and was
experimentally observed about three decades ago
\cite{Fad-Shir-69,Heldman-69,Fad-Shir-70,Carver-72,WHG-73}.
The magnitude of the 3s spectral splitting according to the van Vleck
theorem \cite{Vleck-34} is determined by
	\begin{equation}
	\Delta E_{ex} =\frac{1}{2l +1}(2S + 1)G^2(3s,3d),
	\label{eq:dEex}
	\end{equation}
where $S$ is the total spin of the ground state of the 3d electrons,
$l$ is the orbital quantum number ($l=2$) and $G^2$(3s,3d) is the
Slater exchange integral.
For 3d metal compounds, the calculated 3s splitting is more
than two times larger than the observed one.
It indicates that the observed 3s splitting is not likely to be
due to the spin exchange only.
This fact was explained by the
intra-shell correlation effects between 3s$^1$3p$^6$3d$^n$ and
3s$^2$3p$^4$3d$^{n+1}$ configurations \cite{Bagus-73,Vii-Ohrn-75}.

This scheme does not take into account the effect of a core-hole
screening in the final state of photoemission.
The final-state screening gives satellites in the x-ray photoelectron
core-level spectra whereas "main peaks" usually correspond to final states
with an extra 3d electron in comparison with the ground state.
Veal and Paulikas \cite{Veal-Paul-83} proposed that the 3s splitting
is determined by the exchange interaction in the 3s$^1$3d$^{n+1}$
configuration rather than in the 3s$^1$3d$^n$ configuration.
Kinsinger \etal \cite{Kinsin-90} have shown that the model of Veal
and Paulikas is correct for Ni and Cu compounds but fails for
d-electron numbers of less than~6.

Oh \etal \cite{Oh-Gw-Park-92} discussed the spectral shape of the
3s XPS for Fe and Mn di-halides, taking into account the intra-atomic
configuration interaction in a phenomenological manner.
They proposed that the interpretation of 3s core-level spectra should
be consistent with that of the 2p spectra.
In this case, the 3s splitting reflects the local moment of the ground
state only when the charge-transfer satellite in the 2p core-level
spectra is negligible.

Okada and Kotani \cite{Ok-Kot-92Jpn} theoretically investigated 2p and
3s core-level spectra in late 3d transition metal di-halides and
monoxides in terms of a cluster model.
It was found that the hybridization with high-order charge-transfer
states can reduce the multiplet splitting of a spectrum.
The calculations of the 3s spectra of CrF$_2$, MnF$_2$ FeF$_2$
and Cr$_2$O$_3$ were carried out, and the importance of covalency in
the final states of XPS was noted \cite{Ok-Kot-94Jpn,Uozumi-97}.

In this paper, we present new experimental data on the 3s splitting
in 3d monoxides, Li-substituted oxides and CuGeO$_3$.
The values of the 3s exchange splitting for LaMnO$_3$, SrMnO$_3$,
SrFeO$_3$ and CuFeO$_2$ are present too.
We show that for the late 3d metal oxides
(CuO, CuGeO$_3$, NiO, CoO, FeO)the exchange splitting
can be seen for both, the 3s$^1$3d$^{n+1}$\underline{L} and the 3s$^1$3d$^n$
configuration.
For earlier 3d metal monoxides and Li-substituted oxides,
the value of the exchange splitting correlates well with that predicted for
the 3s$^1$3d$^n$ configuration.
This effect is explained on the basis of a simple configuration-mixing
model.
For LiCoO$_2$ with the 3d$^6$ ground-state configuration ($S=0$) no
exchange splitting is fixed.

\section{Experimental conditions}
\label{Sec.II}

The experiments were performed using a PHI 5600 ci multitechnique system.
${\rm Al}~K_\alpha$ radiation was monochromatized by a double-focusing
monochromator giving a spot diameter of 0.8~mm at the sample position.
Electrons were analyzed from an area of 0.4~mm in diameter.
The energy resolution as determined at the
Fermi level of an Au-foil was 0.3--0.4~eV. All spectra were calibrated
using an Au-foil with $E_B$(4f$_{7/2}) = 84.0$~eV.
All samples were cleaved in vacuo at a base pressure of $5 \times 10^{-10}$
Torr.
For CuO, NiO, CoO, FeO, MnO
and CuGeO$_3$ single crystals have been used.
The lithium-substituted samples (LiCoO$_2$, LiMnO$_2$, Li$_2$MnO$_3$,
LiFeO$_2$ and LiCrO$_2$), manganites (LaMnO$_3$, SrMnO$_3$)
and SrFeO$_3$ have been
prepared using a ceramic technology \cite{GKU-95,Keller-97}.

\section{Results and discussion}
\label{Sec.III}

\begin{figure}
\epsfysize=22.0cm
\hspace*{0.0cm}\epsfbox{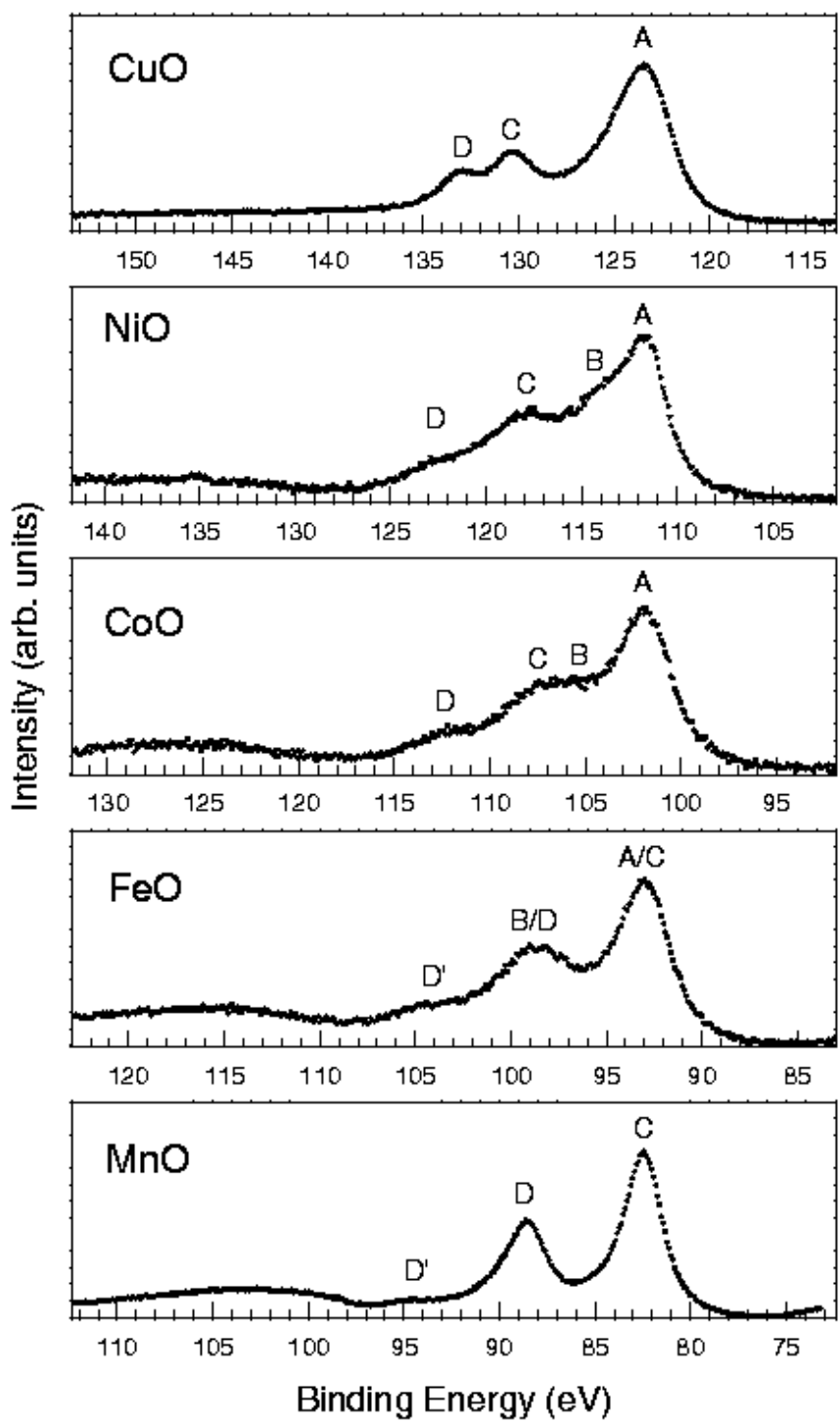}
\caption{3s photoelectron spectra of CuO, NiO, CoO, FeO and MnO.}
\label{f:XPS-Cu-Mn}
\end{figure}

The 3s spectra of the 3d monoxides MnO, FeO, CoO, NiO and CuO are
shown in figure \ref{f:XPS-Cu-Mn}.
The 3s spectrum of MnO shows two sharp peaks, labelled $C$ and
$D$, and a satellite $D'$ at about 6~eV from the peak $D$.
For FeO, the peaks are wider, and for the late metal oxides, CoO
and NiO, the metal 3s spectra exhibit a complex structure (peaks
$A$, $B$, $C$ and $D$).
In the case of CuO, three peaks ($A$, $C$ and $D$) can be distinguished.

\begin{figure}
\epsfysize=20.0cm
\hspace*{2.5cm}\epsfbox{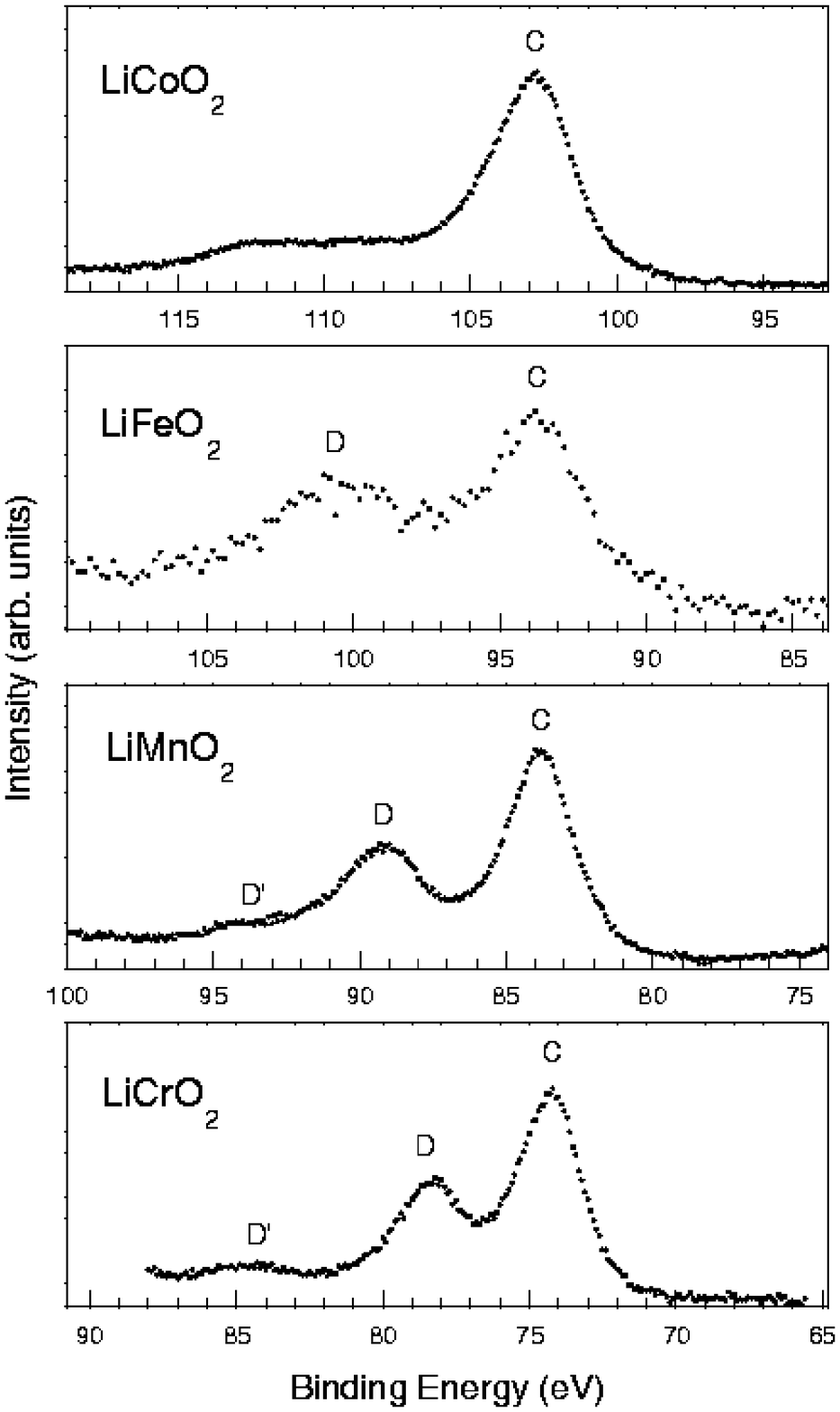}
\caption{3s photoelectron spectra of LiCoO$_2$, LiFeO$_2$, LiMnO$_2$,
        and LiCrO$_2$.}
\label{f:XPS-LiCo-LiCr}
\end{figure}

Figure \ref{f:XPS-LiCo-LiCr} shows the x-ray photoelectron 3s spectra of
lithium-substituted oxides LiCoO$_2$, LiFeO$_2$, LiMnO$_2$ and LiCrO$_2$.
The spectra of LiFeO$_2$, LiMnO$_2$ and LiCrO$_2$ show the peaks $C$ and $D$,
and the satellite $D'$. The spectrum of LiCoO$_2$ shows one sharp peak
at the binding energy 103~eV and the satellite structure extended from
about 107 to 115~eV.

The peaks $C$ and $D$ in the MnO, LiFeO$_2$, LiMnO$_2$ and
LiCrO$_2$ spectra can be explained by the exchange splitting in
the 3s$^1$3d$^n$ final state configurations.
For CoO and NiO, four peaks ar due to the exchange splitting
of 3s$^1$3d$^n$ and 3s$^1$3d$^{n+1}$\underline{L} final-state configurations
with $n=8$ for NiO and $n=7$ for CoO.
One can suggest that the peaks $A$ and $B$ arise mainly from exchange
splitting in the 3d$^1$3d$^{n+1}$\underline{L} configurations, and the peaks
$C$ and $D$ belong to the 3s$^1$3d$^n$ configurations.
In the case of FeO we cannot attribute a peak to a certain configuration,
since each of the peaks is a mixture of different configurations.

For CuO, the exchange splitting can be expected for the 3s$^1$3d$^9$
configuration only. Consequently three peaks, marked $A$, $C$ and $D$,
are found.
The peaks $C$ and $D$ at energies 130 and 133~eV are high-spin and
low-spin states of the 3s$^1$3d$^9$ configuration, respectively.
A single peak $A$ with the binding energy 123.4~eV arises due to the
3s$^1$3d$^{10}$\underline{L} final states.

The Co$^{3+}$ ground state of LiCoO$_2$ can be written as a low-spin
t$_{2g \uparrow}^3$t$_{2g \downarrow}^3$ state with $S=0$.
The non-magnetic character of Co$^{3+}$ ions in LiCoO$_2$ has been found
by Bongers \cite{Bongers}.
In this situation no exchange splitting should be expected for the
Co~3s spectra in LiCoO$_2$.

\begin{figure}
\epsfysize=11.0cm
\hspace*{2.0cm}\epsfbox{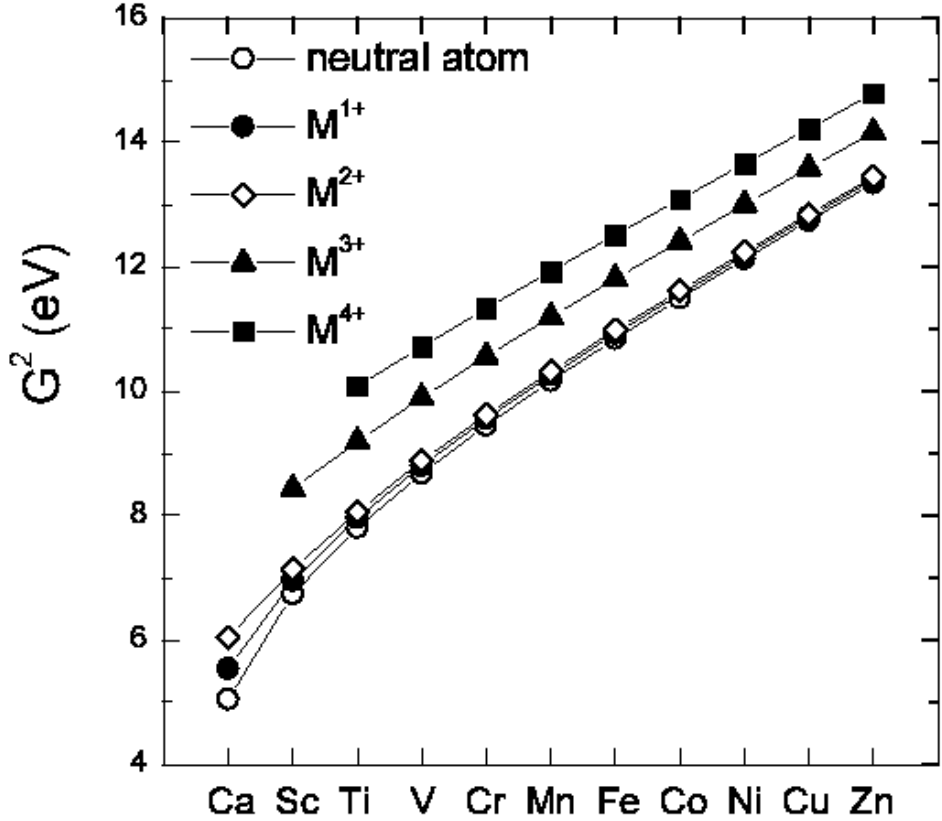}
\caption{Slater exchange integrals for neutral atoms, monovalent, divalent,
         and trivalent ions of 3d metals (eV).}
\label{f:Slater}
\end{figure}

\begin{figure}[th]
\epsfxsize=14.0cm
\hspace*{2.0cm}\epsfbox{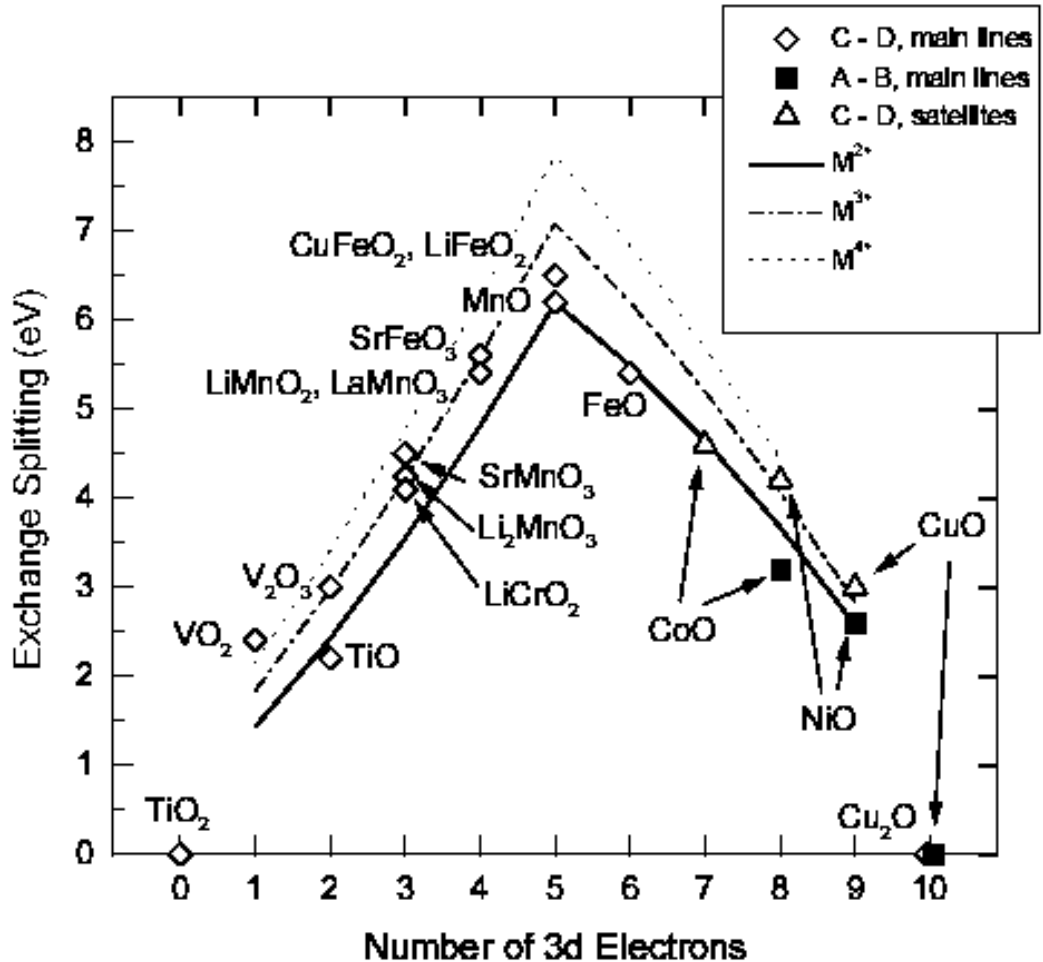}
\caption{Exchange splitting of different transition-metal oxides plotted as
         a function of the number of 3d electrons in the ground state.
         Full triangles represent the splitting determined from the $C$--$D$
         lines for early transition-metal oxides.
         Open triangles represent the splitting determined from the $C$--$D$
         lines for late metal oxides.
         Full diamonds represent the splitting determined from the $A$--$B$
         lines for the late transition-metal oxides.
         For the $A$--$B$ lines, the number of 3d electrons is increased
         by one.
         The lines give the exchange splitting calculated  for divalent,
         trivalent and tetravalent ions.
         The calculated values are reduced by 50\% to be compared with the
         experimental values.}
\label{f:Split-Z}
\end{figure}

According to equation (\ref{eq:dEex}), the 3s energy splitting depends
on both, the Slater exchange integral $G^2$(3s,3d) and the total spin S
of the 3d electrons in the ground state.
Figure~\ref{f:Slater} shows the Slater exchange integral calculated for free
neutral atoms, monovalent, divalent, trivalent and tetravalent ions.
Note that for $S=0$, the size of the splitting is not determined by
equation~(\ref{eq:dEex}), since the energies for both final states are
equal to zero:
\begin{equation}
E_{ex}^{S-\frac{1}{2}} = E_{ex}^{S+\frac{1}{2}}=0.
\end{equation}
The calculated splitting of the 3s states for 3d transition-metal
ions is much larger than observed.
Bagus \etal \cite{Bagus-73} explained the small measured splitting
by an interaction between the 3s$^1$3p$^6$3d$^n$ and 3s$^2$3p$^4$3d$^{n+1}$
configurations along with the electron correlation in the final core-hole
states.
In order to compare the calculated with the experimental energy splitting,
we reduced the calculated value by about 50\%, based on the comparison of
the calculated splitting for a neutral Mn atom with the measured one for
Mn in the gas phase \cite{Fadley-88}.
The measured value of the Mn~3s splitting for Mn atoms is 6.5~eV whereas
the calculated one is 12.2~eV.

\begin{figure}
\epsfxsize=12.0cm
\hspace*{3.0cm}\epsfbox{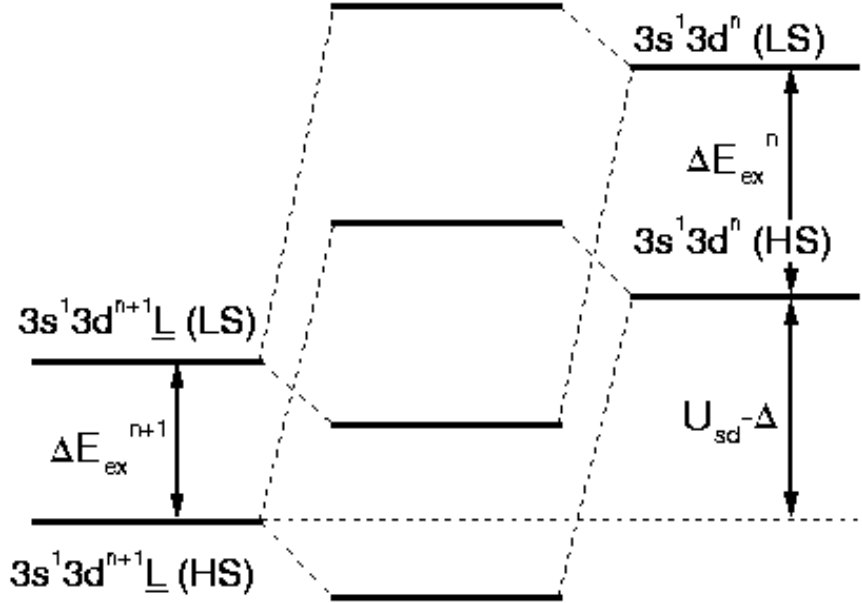}
\caption{Schematic energy diagram showing the configuration mixing.
         The multiplet splitting in the 3s$^1$3d$^{n}$ state is reduced
         by the mixing with the 3s$^1$3d$^{n+1}$\underline{L} state.}
\label{f:Diagram}
\end{figure}

In figure~\ref{f:Split-Z} we show the measured exchange splitting of the
3s core levels, as a function of the 3d electrons in the ground state.
The lines give the magnitude of the spectral splitting calculated
for divalent, trivalent and tetravalent ions, reduced by 50~\%.
The experimental data present compounds with monovalent (Cu$_2$O),
divalent (TiO, MnO, FeO, CoO, NiO and CuO), trivalent (V$_2$O$_3$,
LiCrO$_2$, LiMnO$_2$, LaMnO$_3$, LiFeO$_2$, CuFeO$_2$ \cite{GPK-97}) and
tetravalent (TiO$_2$ and VO$_2$) 3d ions.
The value for LiCoO$_2$ is not shown here, since the ground state of
this oxide is characterized by the low-spin configuration $S=0$,
whereas figure~\ref{f:Split-Z} is plotted for the oxides in the high-spin
configurations.

The experimental data are divided into three groups.
In the first group data are plotted for the early 3d monoxides
(from TiO to MnO) and for some complex oxides (LiMnO$_2$, Li$_2$MnO$_3$,
LaMnO$_3$, LiCrO$_2$, LiFeO$_2$, CuFeO$_2$, SrMnO$_3$, SrFeO$_3$) based
on the measurements of the spectral splitting between the $C$
and $D$ components.
In the second group data are represented for the late 3d oxides (CoO,
NiO and CuO) determined from the measured $C$--$D$ spectral splitting.
In the third group data are included deduced from the measured $A$--$B$
lines for CoO, NiO and CuO.
The latter data are plotted for 3d electron numbers increased by one.
It is assumed that for the late transition-metal oxides the
3d$^{n+1}$\underline{L} ground state configuration is realized.
For FeO, the estimated value of 5.4~eV is tentatively assigned to the first
group data (spectral splitting between $C$ and $D$ peaks).

In this simple scheme it is assumed that the main 3s XPS peaks of the early
3d oxides originate from a 3s$^1$3d$^n$ final-state configuration, and
those of the late 3d oxides from a 3s$^1$3d$^{n+1}$\underline{L}
final-state configuration.
In reality, both the 3s$^1$3d$^n$ and 3s$^1$3d$^{n+1}$\underline{L} state are
present in the 3s spectra of both, the early and late 3d oxides.
The quantitative estimation of the contributions of both configurations can
be made on the basis of a model including the mixing of the 3s$^1$3d$^n$ and
3s$^1$3d$^{n+1}$\underline{L} final-state configurations.
We demonstrate this in figure~\ref{f:Diagram} for a model in which we
consider two (high-spin and low-spin) 3s$^1$3d$^n$ states and two
3s$^1$3d$^{n+1}$\underline{L} states.
Each of these states with the energy $E^{HS}$ or $E^{LS}$ mixes with the
other with the same spin, this leads to the reduction of the multiplet
splitting depending on the covalent mixing in the final state.
This model is a two-level model employed separately to high-spin and
low-spin states.

\begin{figure}
\epsfysize=20.0cm
\hspace*{1.5cm}\epsfbox{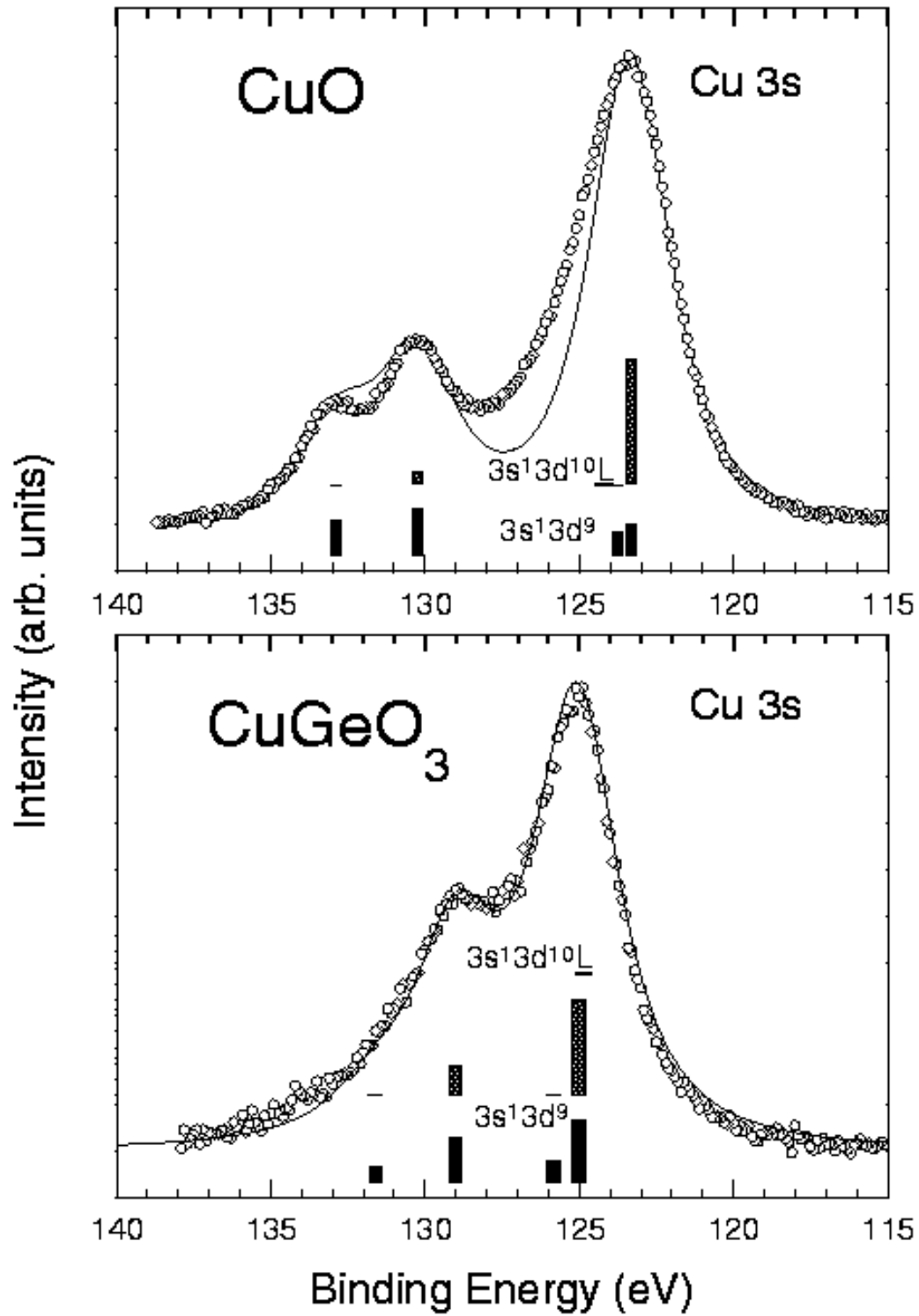}
\caption{Experimental and theoretical Cu~3s  spectra for CuO  and
         CuGeO$_3$.
         In the lower parts of the figures the contributions of the
         3s$^1$3d$^9$ and 3s$^1$3d$^{10}$\underline{L} final-state
         configurations are shown.}
\label{f:CuO-CuGeO3}
\end{figure}

\begin{figure}
\epsfysize=20.0cm
\hspace*{1.5cm}\epsfbox{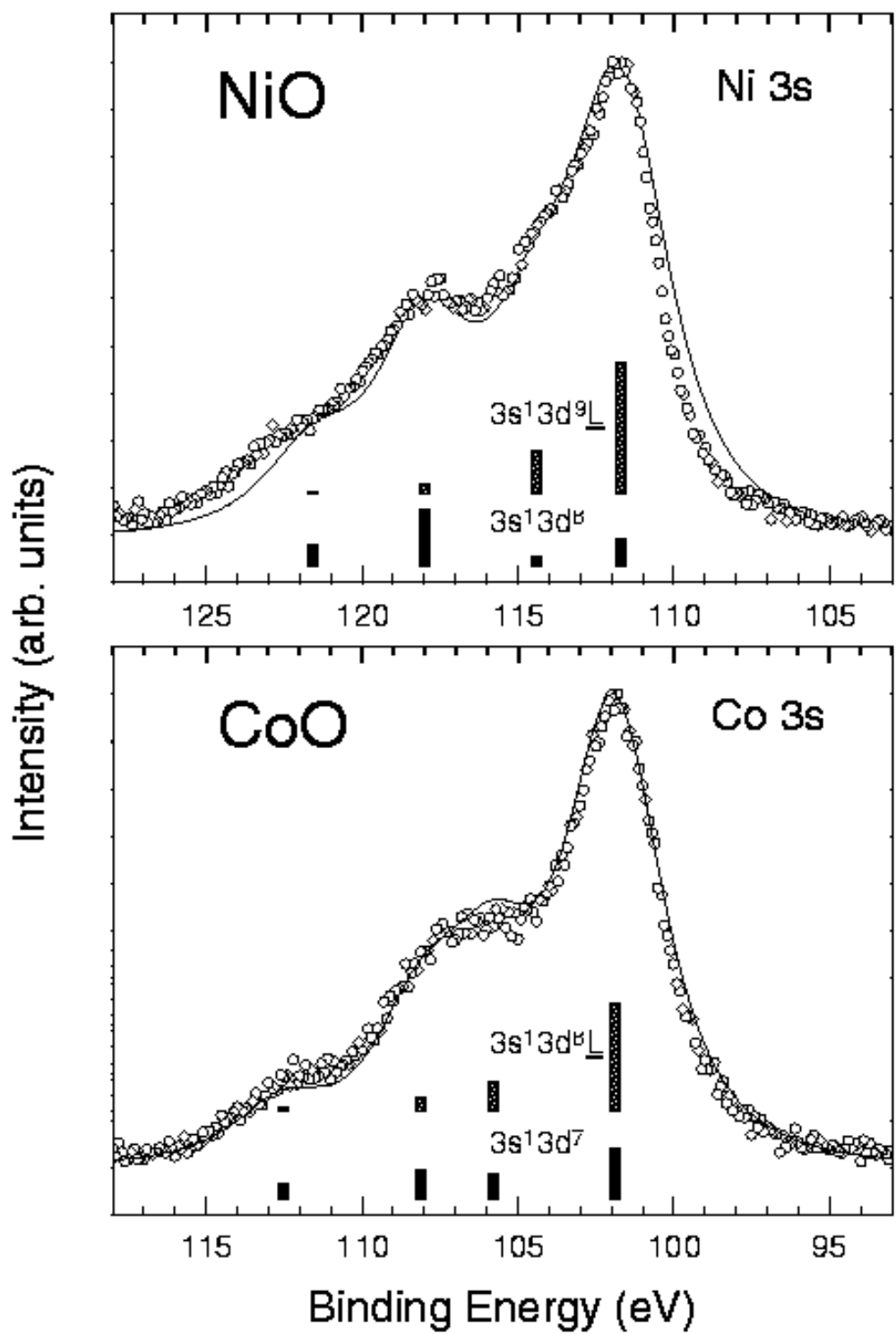}
\caption{Experimental and theoretical Ni~3s and Co~3s spectra for NiO
         and CoO.
         In the lower parts of the figures the contributions of the
         3s$^1$3d$^8$ and 3s$^1$3d$^9$\underline{L} (NiO) and 3s$^1$3d$^7$
         and 3s$^1$3d$^8$\underline{L} (CoO) final-state configurations
         are shown.}
\label{f:NiO-CoO}
\end{figure}

\begin{figure}
\epsfysize=20.0cm
\hspace*{1.5cm}\epsfbox{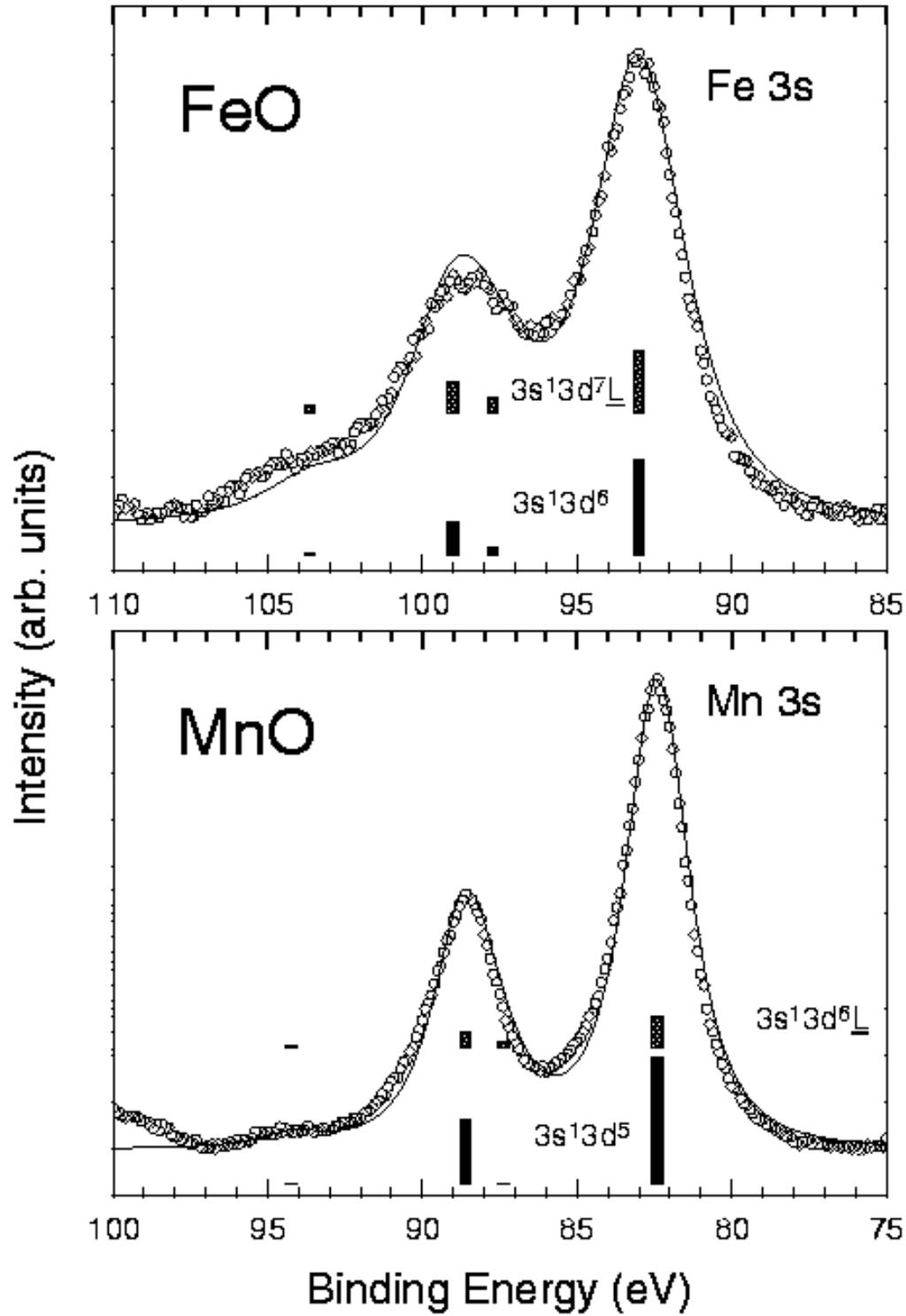}
\caption{Experimental and theoretical Fe~3s and Mn~3s spectra for FeO
         and MnO.
         In the lower parts of the figures the contributions of the
         3s$^1$3d$^6$ and 3s$^1$3d$^7$\underline{L} (FeO) and 3s$^1$3d$^5$
         and 3s$^1$3d$^6$\underline{L} (MnO) final-state configurations
         are shown.}
\label{f:FeO-MnO}
\end{figure}

\begin{figure}
\epsfysize=20.0cm
\hspace*{1.5cm}\epsfbox{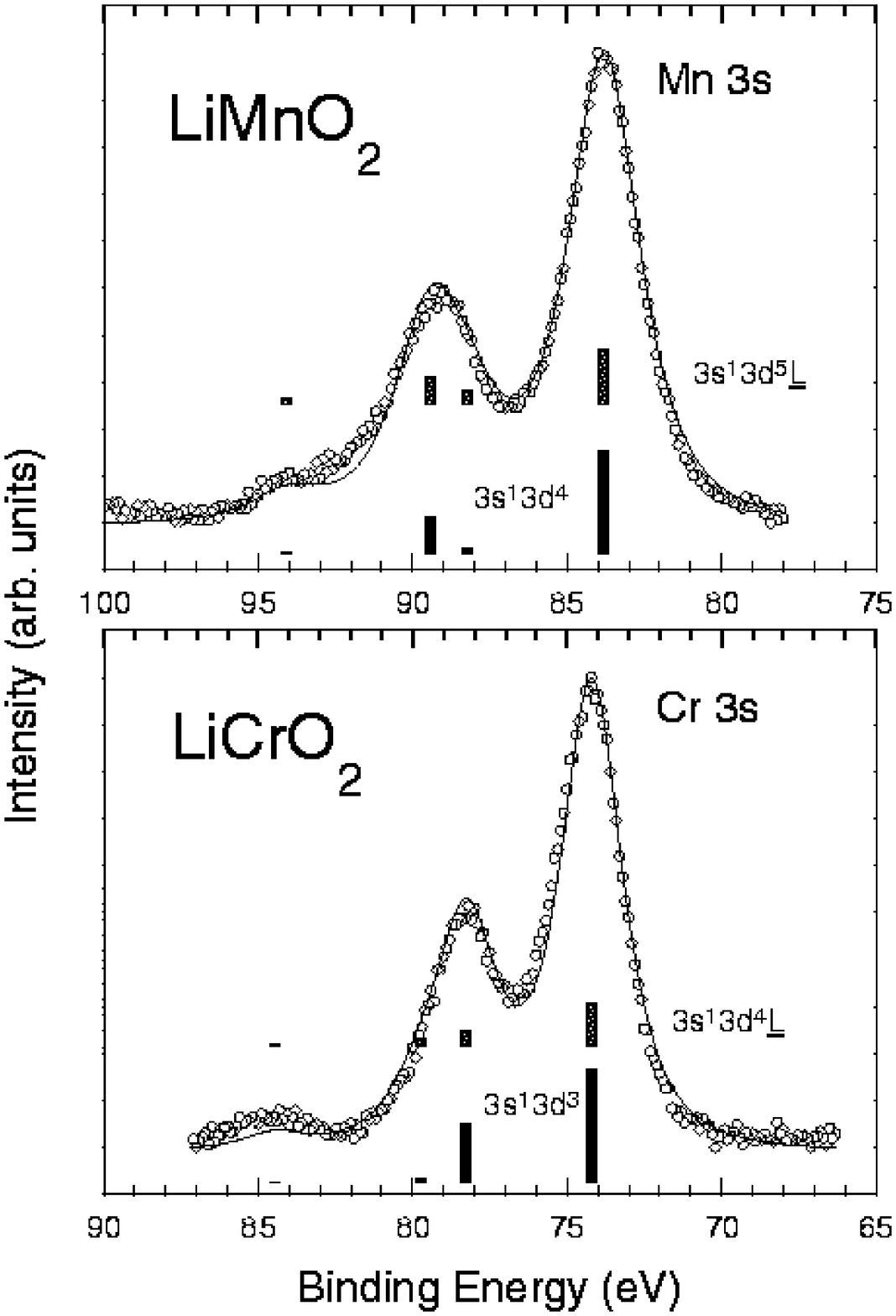}
\caption{Experimental and theoretical Mn~3s and Cr~3s spectra for
        LiMnO$_2$ and LiCrO$_2$.
        In the lower parts of the figures the contributions of the
        3s$^1$3d$^4$ and 3s$^1$3d$^5$\underline{L} (LiMnO$_2$) and
        3s$^1$3d$^3$ and 3s$^1$3d$^4$\underline{L} (LiCrO$_2$)
        final-state configurations are shown.}
\label{f:LiMn-LiCr}
\end{figure}

The calculated 3s spectra for CuO, CuGeO$_3$, NiO, CoO, FeO, MnO,
LiMnO$_2$ and LiCrO$_2$ are shown in the
figures~\ref{f:CuO-CuGeO3}, \ref{f:NiO-CoO}, \ref{f:FeO-MnO} and
\ref{f:LiMn-LiCr}
together with the experimental data. The experimental spectra are corrected
for the background arising due to secondary electrons.
The parameters $U_{sd}$ (the core-hole d electron Coulomb attraction
energy), $\Delta$ (the charge transfer energy) and $T$ (the transfer
integral) used for  the calculations are given in the table.
The lower parts of the figures show
3s$^1$3d$^n$ and 3s$^1$3d$^{n+1}$\underline{L} contributions forming the
main line and satellites, both in high-spin and low-spin configurations.

\begin{table*}[th]
\label{Tab-1}
\caption{Parameters used to calculate the 3s-XPS spectra for 3d metal
         oxides and contributions of the 3d$^{n+1}$\underline{L}
         ($\alpha_{1}^{2}$) and 3s$^1$3d$^{n+1}$\underline{L}
         ($\beta_{1}^{2}$) configurations to the ground and final states,
         respectively (for explanation see text).}
\begin{indented}
\item[]
\begin{tabular*}{13.0cm}{@{}l@{\extracolsep{\fill}}ccccc}
\br
 & $U_{sd}$ (eV) & $\Delta$ (eV) & $T$ (eV) & $\alpha_1^2$ & $\beta_1^2$ \\
\mr
CuO        & 7.0  & 2.9    & 2.8    &   0.27  &  0.80 \\
CuGeO$_3$  & 7.0  & 6.0    & 2.0    &   0.08  &  0.62 \\
NiO        & 6.5  & 2.4    & 2.4    &   0.28  &  0.83 \\
CoO        & 6.0  & 4.0    & 2.9    &   0.22  &  0.68 \\
FeO        & 5.8  & 6.8    & 2.3    &   0.09  &  0.39 \\
MnO        & 5.0  & 8.0    & 2.0    &   0.05  &  0.20 \\
LiMnO$_2$  & 5.5  & 6.9    & 2.1    &   0.07  &  0.34 \\
LiCrO$_2$  & 5.0  & 7.9    & 2.5    &   0.08  &  0.28 \\
\br
\end{tabular*}
\end{indented}
\end{table*}

One can see that the large parameter $\Delta$ for CuGeO$_3$ leads to a
smaller distance between the main line and the satellite in comparison to
CuO, which has the same valence state of copper ions (Cu$^{2+}$)
\cite{Parmigiani-97-CuGeO3}.
It is obvious, that the magnitude of the exchange splitting depends on the
hybridization parameters.
The large value $\Delta$ reflects the strong ionic character of CuGeO$_3$
with respect to CuO.
Note, that in Ref.~\cite{Parmigiani-97-CuGeO3} the parameters $\Delta$ were
determined from Cu~2p core-level spectra on the basis of the Anderson
Hamiltonian model in the impurity limit for CuO and CuGeO$_3$ as 1.75 and
4.2~eV, respectively.
They are lower than those calculated in our work.
On the other hand, the value $\Delta$ for CuO is in the range from 1.15~eV
to 3.5~eV \cite{Tjeng-CuO,Parlebas-93-CuO,Veen-93-CuO,Zimmer-PhD},
and the parameter $\Delta$ for CuO, determined in our work, corresponds
closely to these data.

The parameter $\Delta$ increases on going from NiO to MnO.
One can see that the main line of the Ni~3s spectrum for NiO
is determined mainly by Ni~3s$^1$3d$^9$\underline{L} final states.
For FeO, the contribution of 3s$^1$3d$^7$\underline{L} and
3s$^1$3d$^6$ states are of about the same value, and for
MnO, the main peak is determined by 3s$^1$3d$^5$ states.
For LiMnO$_2$ the parameter $\Delta$ is about 6.9~eV, similar to that
of FeO.
The LiCrO$_2$ oxide is more ionic than LiMnO$_2$, as can be seen from
the comparison of the values $\Delta$.

For the LiCoO$_2$, the satellite near the strong line can be explained
by the exchange splitting of the s$^1$d$^7$\underline{L} final-state
configuration. Note, oxygen vacancies due to non-stoichiometry in
LiCoO$_{2-x}$ can lead to an appearance of Co$^{2+}$ ions, which should be
visible in the 3s spectrum.

This simple two-level model allows to explain the
3s splitting of transition-metal oxides as a function of the number
of 3d electrons in the ground state, as presented in figure~\ref{f:Split-Z}.
For the late 3d oxides, the final-state configuration is determined mainly
by 3s$^1$3d$^{n+1}$\underline{L} well-screened states
($\beta ^2 > 0.5$). Therefore, the 3s spectra of the late oxides exhibit
four peaks (or three peaks, as in the case of CuO).
The final-state configurations of
the early 3d oxides (the monoxides and complex oxides with trivalent and
tetravalent 3d ions) are decribed mainly by poorly screened 3s$^1$3d$^n$
states. In this case, the spectra show two sharp peaks and a satellite.
The satellite in the 3s spectra can be explained on the basis of
the presented model.

We demonstrated  the possibility to extract additional information about
intra- and interatomic exchange effects from 3s core-level x-ray
photoelectron spectra.
One can see that the interaction between the high-spin and low-spin
3s$^1$3d$^n$ and 3s$^1$3d$^{n+1}$\underline{L} final-state configurations
can lead to radical changes of the exchange splitting, in comparison with
that predicted by the van Vleck theorem.
Although the presented model does not take into account crystal-field
effects and 3s$^1$3d$^{n+2}$\underline{L}$^2$ configurations, it can explain
all the  experimental facts concerning 3s core-level spectra of 3d
transition-metal oxides.

\section{Conclusion}
\label{Sec.IV}

We have presented new experimental data on the 3s metal spectra of 3d
transition-metal oxides.
The 3s splitting of the 3d oxides can be well
defined as a function of the 3d electron number in the ground state
both for d$^n$ and d$^{n+1}$\underline{L} configurations.
It was  shown that the
spectra can be analysed using the simple two-configuration
model of the interatomic configuration mixing.
The change of the 3s spectra is ascribed to the change of the
charge-transfer energy.

\ack
This work was supported by the DFG-RFBR Rroject,
the NATO Project (Grant No.~HTECH.LG940861),
the International Project "Electron Structure of Oxides",
the Russian Foundation for Fundamental Research (Grant No.~96-15-96598),
and the Deutsche Forschungsgemeinschaft through the Graduiertenkolleg
"Mikrostruktur oxidischer Kristalle" and the SFB 225.

\section*{Appendix}
\appendix

The ground-state wave function can be written in the form
	\begin{equation}
|g\rangle = \alpha_0 |d^n \rangle -  \alpha_1 |d^{n+1}\underline L \rangle,
	\end{equation}
where $\alpha_0^2 + \alpha_1^2 = 1$.
For the high spin-state configuration, the final-state wave
function is written as
	\begin{eqnarray}
	|f_1\rangle = \beta_0 |s^1 d^n \rangle -
	\beta_1 |s^1 d^{n+1}\underline L \rangle,
	\nonumber \\
	|f_2\rangle = \beta_1 |s^1 d^n \rangle +
	\beta_0 |s^1 d^{n+1}\underline L \rangle,
	\end{eqnarray}
and  the eigenvalues are
	\begin{equation}
	E_{1,2}^{HS} = \frac{1}{2} (U_{sd}-\Delta) \pm \frac{1}{2}
	\sqrt{(U_{sd}-\Delta)^2 + 4 T^2}.
	\end{equation}

Here, $\Delta$ is the charge transfer energy, $U_{sd}$ is the
core-hole $d$ electron Coulomb attraction energy, and $T$ is the transfer
integral, defined as follows:
	\begin{eqnarray}
	& \langle s^1 d^n | \hat H | s^1 d^n \rangle  =  U_{sd} - \Delta,
	\nonumber \\
	& \langle s^1 d^{n+1} \underline L | \hat H |
	s^1 d^{n+1} \underline L \rangle  =  0, \\
	& \langle s^1 d^n | \hat H | s^1 d^{n+1} \underline L \rangle  = T.
	\nonumber
	\end{eqnarray}

Let us define $\Delta E_{ex}^{n}$ as the exchange energy for
the $s^1 d^n $ final states and $\Delta E_{ex}^{n+1}$
as the exchange energy for the $s^1 d^{n+1} \underline L$ states.
In this case, the eigenvalues for the low-spin configuration are:
	\begin{eqnarray}
	E_{3,4}^{LS} = \Delta E_{ex}^{n+1} & + & \frac{1}{2}
\left[(U_{sd} - \Delta) + (\Delta E_{ex}^n - \Delta E_{ex}^{n+1}) \right ]
	\pm
	\nonumber \\
	&\pm & \frac{1}{2}
	\sqrt{\left[(U_{sd} - \Delta) + (\Delta E_{ex}^n -
	\Delta E_{ex}^{n+1}) \right ]^2 + 4 T^2 }.
	\end{eqnarray}
Coefficients $\alpha_0$, $\alpha_1$, $\beta_0$, $\beta_1$
are determined as follows:
	$\alpha_0 = \cos \theta_0 $,
	$\alpha_1 = \sin \theta_0 $,
	$\beta_0  = \cos \theta_1 $ and
	$\beta_1  = \sin \theta_1 $, where
	\begin{eqnarray}
	\tan(2\theta_0) & = & \frac{2T}{\Delta}, \\
	\tan(2\theta_1) & = & \frac{2T}{\Delta - U_{sd}}.
	\end{eqnarray}

For the high-spin configuration, the ratio of the intensity of the
satellite to the main line is given (in the sudden approximation) by:
	\begin{equation}
	\frac{I_{2}^{HS}} {I_{1}^{HS}} =
	\frac{\left | \langle f_2|g\rangle \right |^2}
	{\left | \langle f_1|g \rangle \right |^2} =
	\frac{(\alpha_0 \beta_1 -\alpha_1 \beta_0)^2}
	{(\alpha_0 \beta_0 + \alpha_1 \beta_1)^2}.
	\end{equation}

In order to calculate the intensities of the lines in the low-spin
configuration, it is necessary to determine the contributions of both, the
$s^1 d^n$ and $s^1 d^{n+1} \underline L$ configurations forming the main
line and the satellite.
The intensity of the main line for the low-spin state is
	\begin{equation}
  I_3^{LS} = I_1^{HS} \left(k^{n+1} \beta_1^2 \frac{S^{n+1}}{S^{n+1}+1} +
	k^{n} \beta_0^2 \frac{S^n}{S^n +1}\right),
	\end{equation}
and for the satellite
	\begin{equation}
  I_4^{LS} = I_2^{HS} \left(k^{n+1} \beta_0^2 \frac{S^{n+1}}{S^{n+1}+1} +
       k^{n} \beta_1^2 \frac{S^n}{S^n +1}\right).
	\end{equation}
Here, $S^n$ and $S^{n+1}$ are the values of the $d$ shell in the $d^n$ and
the $d^{n+1}\underline L$ configurations, respectively.
To take into account the $3s^1 3p^6 3d^n$~-- $3s^2 3p^4 3d^{n+1}$
and  $3s^1 3p^6 3d^{n+1}$~-- $3s^2 3p^4 3d^{n+2}$
configuration interactions, we used coefficients $k^n$ and $k^{n+1}$,
respectively.
Viinikka and \"Ohrn \cite{Vii-Ohrn-75} showed that the configuration
interaction leads to a splitting of the low-spin term into new lines
at energies approximately $20$--$45$ eV separated from the high-energy
line.
In consequence of this, the relative intensity of the  main low-spin peak is
lower than expected from the simple multiplicity ratio $\frac{S}{S+1}$.
For example, the calculated intensity ratio for the $^5S$ to $^7S$ states
for Mn$^{2+}$ ions is equal to 0.47 \cite{Vii-Ohrn-75} which is lower than
$\frac{S}{S+1}=0.71$.
The values $k^n$ and $k^{n+1}$ for MnO and CuO were determined from a
comparison of experimental and calculated spectra, and for NiO, CoO and FeO
they were estimated according to Ref.~\cite{Vii-Ohrn-75}
(see table~\ref{Tab-1}).
The parameters $U_{sd}$ for divalent compounds were taken from the paper
of Okada and Kotani \cite{Ok-Kot-92Jpn}.
For trivalent compounds, LiMnO$_2$ and LiCrO$_2$, the values $U_{sd}$ were
taken according to \cite{Uozumi-97}.
We assumed that $U_{sd} = U_{pd} - 1$~eV, where $U_{pd}$ is is the
$2p - 3d$ electron Coulomb attraction energy.

The coefficient $k^n$ for CuO was taken as 2.2.
This value cannot be explained by the interaction between the
$3s^1 3p^6 3d^9$~-- $3s^2 3p^4 3d^{10}$ configurations, since this
interaction should lead to a decrease of the low-spin component intensity.
The enhanced intensity of the peak at 133~eV can be explained if one
takes into account crystal-field effects.
The crystal field should lead to both, to an increase of the intensity of the
low-spin peak $D$ and to an asymmetry of the main peak $A$.

Note that the intensity of the $^7S$ term in the Mn~$3s$ spectrum of MnO
increases due to a Mn~$3p$ shake satellite, which is situated at the same
energy \cite{StUhl}.
This leads to a virtual decrease of the relative intensity of the low-spin
peak.

The solid curves are obtained by a convolution of the calculated line
spectrum with the function
	\begin{equation}
    I(E) = \frac {I_0 \gamma ^3}{\left [(E-E_0)^2+\gamma^2 \right] ^{1.5}},
	\end{equation}
where $E_0$ is the energy  of the lines corresponding to each of the
configuration contribution, and $I_0$ is the intensity of the lines.
The parameter $\gamma$ is a value for the convolution.
We have chosen this function for the convolution as it is an intermediate
between the Lorentzian and Gaussian functions.


\section*{References}
\begin{thebibliography}{99}

\bibitem{Fad-Shir-69} Fadley C S, Shirley D A, Freeman A J,
	Bagus P S and Mallow J V
        1969 {\it Phys. Rev. Lett.} {\bf 23} 1397
\bibitem{Heldman-69} Heldman J, Heden P F, Nordling C and Siegbahn K
        1969 {\it Phys. Lett.} A {\bf 29} 178
\bibitem{Fad-Shir-70} Fadley C S and Shirley D A
        1970 {\it Phys. Rev.} A {\bf 2}	1109
\bibitem{Carver-72} Carver J C, Schweitzer G K and Carlson T A
	1972 {\it J. Chem. Phys.} {\bf 57} 973
\bibitem{WHG-73} Wertheim G K, H\"ufner S and Guggenheim H J
        1973 {\it Phys.	Rev.} B {\bf 7} 556
\bibitem{Vleck-34} van Vleck J H 1934 {\it Phys. Rev.} {\bf 45} 405
\bibitem{Bagus-73} Bagus P S, Freeman A J and Sasaki F
	1973 {\it Phys. Rev. Lett.} {\bf 30} 850
\bibitem{Vii-Ohrn-75} Viinikka E-K and \"Ohrn Y
        1975 {\it Phys. Rev.} B {\bf 11} 4168
\bibitem{Veal-Paul-83} Veal B W and Paulikas A P
        1983 {\it Phys. Rev. Lett.} {\bf 51} 1995
\bibitem{Kinsin-90} Kinsinger V, Sander I, Steiner P, Zimmermann R
	and H\"ufner S 1990 {\it Solid State Commun.} {\bf 73} 527
\bibitem{Oh-Gw-Park-92} Oh S J, Gweon G H and Park J G
        1992 {\it Phys. Rev. Lett.} {\bf 68} 2850
\bibitem{Ok-Kot-92Jpn} Okada K and Kotani A
        1992 {\it J. Phys. Soc. Japan} {\bf 61} 4619
\bibitem{Ok-Kot-94Jpn} Okada K, Kotani A, Kinsinger V, Zimmermann R
	and H\"ufner S 1994 {\it J. Phys. Soc. Japan} {\bf 63} 2410
\bibitem{Uozumi-97} Uozumi T, Okada K, Kotani A, Zimmermann R,
	Steiner P, H\"ufner S, Tezuka Y and Shin S
        1997 {\it J. Electron Spectrosc. Relat. Phenom.} {\bf 83} 9
\bibitem{GKU-95} Galakhov V R, Kurmaev E Z, Uhlenbrock S, Neumann M,
	Kellerman D G and Gorshkov V S
        1995 {\it Solid State Commun.} {\bf 95}	347
\bibitem{Keller-97} Kellerman D G, Gorshkov V S, Zubkov V G,
	Perelyaev V A, Galakhov V R, Kurmaev E Z, Uhlenbrock S and
	Neumann M 1997 {\it Russian J. Inorganic Chemistry} {\bf 42} 1012
\bibitem{Fadley-88} Fadley C S 1988 {\it Core-Level Spectroscopy in
	Condensed Systems} ed J Kanamori and A Kotani
	(Berlin: Springer) p 236
\bibitem{GPK-97} Galakhov V R, Poteryaev A I, Kurmaev E Z, Anisimov V I,
                 Bartkowski S, Neumann M, Lu Z W, Klein B M and
                 Zhao Tong-Rong 1997 {\it Phys. Rev.} B {\bf 56} 4584
\bibitem{Bongers} Bongers P F 1957 PhD Thesis, University of Leiden,
        Leiden, The Netherlands
\bibitem{StUhl} Uhlenbrock S, Mayer B and Neumann M ({\it to be published}).
\bibitem{Parmigiani-97-CuGeO3} Parmigiani F, Sangaletti L, Goldoni A,
        del~Pennino U, Kim C, Shen Z-X, Revcolevschi A and Dhal\`{e}nne G
        1997 {\it Phys. Rev.} B {\bf 55} 1459
\bibitem{Tjeng-CuO} Tjeng L H 1990 PhD Thesis, University of Groningen,
        Groningen, The Netherlands
\bibitem{Parlebas-93-CuO} Parlebas J C
        1993 {\it Phys. Status Solidi} (b) {\bf 178} 9
\bibitem{Veen-93-CuO} van Veenendaal M A and Sawatzky G A
        1993 {\it Phys. Rev. Lett.} {\bf 70} 2459
\bibitem{Zimmer-PhD} Zimmermann R 1996 PhD Thesis, Universit\"at des
        Saarlandes, Saarbr\"ucken, Germany

\end{thebibliography}
\end{document}